\documentstyle[12pt]{article}

\newcommand{\be}{\begin{equation}}
\newcommand{\ee}{\end{equation}}

\newcommand{\bea}{\begin{eqnarray}}
\newcommand{\eea}{\end{eqnarray}}

\newcommand{\p}{\partial}

\newcommand{\nn}{\nonumber \\}
\newcommand{\f}{\frac}

\begin{document}
\thispagestyle{empty}

\begin{flushright}
{\bf arXiv: 0910.0101}
\end{flushright}
\begin{center} \noindent \Large \bf
$\eta/s$ at finite coupling
\end{center}

\bigskip\bigskip\bigskip
\vskip 0.5cm
\begin{center}
{ \normalsize \bf  Shesansu Sekhar Pal}

\vskip 0.5cm

\vskip 0.5 cm
Center for Quantum Spacetime, \\
Sogang University, 121-742 Seoul, South Korea\\
\vskip 0.5 cm
\sf shesansu${\frame{\shortstack{AT}}}$gmail.com
\end{center}
\centerline{\bf \small Abstract}

We compute the ratio of the coefficient of shear viscosity to
entropy density at finite coupling and at zero chemical potential
using holographic duality, up to ten derivative terms in the low
energy effective 5-dimensional action, of a specific kind, which may or may not be connected to the supersymmetric completion of Type IIB theory. The result suggests that this ratio can be
positive only for the 8th derivative term even with the form  of
that term in the action as ${{\cal C}^{ij}}_{kl}{{\cal
C}^{kl}}_{mn}{{\cal C}^{mn}}_{rs}{{\cal C}^{rs}}_{ij}$, where ${\cal
C}$ is the Weyl tensor.\\
\begin{flushleft}
PACS: 11.25.Tq, 11.25.Wx, 11.25.-w, 11.25.Sq
\end{flushleft}
\newpage

\section{Introduction}
It has become very interesting to understand the properties of the newly discovered state of matter, which is
characterized by having very high
energy density, strong collective flow and early thermalization time \cite{expts},  at RHIC and may be called as
the "first signature" for the formation of quark-gluon plasma. However, the strong criteria to see such a phase
would be the quark deconfinement and chiral symmetry restoration  \cite{expts}.
Moreover, since the phase is seen at the strongly coupled limit means the quantitative study of this phase is going to be
very cumbersome. However, we can understand the properties of some other theoretically constructed phases of similar
kind, which may share some of its properties in common with sQGP. Let us  assume that is the case and denote
this theoretically constructed phase  as hQGP. This means  some universal properties could exist which are in  common to both sQGP as well as to hQGP, i.e. those
properties that remain  the same  irrespective of the model understudy. Now, we can understand quantitatively the
properties of these  phase using the holographic prescription \cite{jm}.

One such quantity is the ratio of shear viscosity to entropy density, $\eta/s$. This ratio has
been computed earlier for ${\cal N}=4$ system at finite temperature \cite{pss} and is found to have the value $\f{1}{4\pi}$.
It has been conjectured in \cite{kss} that for any (relativistic) system that admits a gravity dual should have a
minimum value  of $\f{1}{4\pi}$ at zero chemical potential. This conjecture was based mainly on the calculations done
(during that time) in a very specific limit that is in the large rank of the gauge group and for large value of
the 't Hooft coupling ($N\rightarrow \infty,~~~\lambda\rightarrow \infty$) and also  in the absence of enough experimental data.
However, recently it has been shown explicitly that for some theoretically constructed models  this ratio can even
go below $\f{1}{4\pi}$ and this happens at finite 't Hooft coupling limit \cite{blmsy} and \cite{kp}. In \cite{td}, a theoretical
argument was provided with the help of a Gedanken type experiments to falsify the KSS conjecture. See \cite{ss},  for
nice reviews on the subject.

In this note, we shall present  results based on the  calculations done
using holographic duality. We evaluate the ratio of $\eta/s$ up to tenth orders in derivative expansion using
the elegant approach
proposed by  Iqbal and Liu \cite{liu} and is beautifully presented to study some more examples in \cite{mps}. In this 
approach \cite{liu}, one important property is: the transport properties at the boundary is fully captured at the horizon
but only in the zero frequency limit. This approach has the drawback, we can evaluate the transport quantities only in the
scalar channel.

In our computation, the only degrees of freedom we have  considered is the metric. It is assumed not to couple with any other
degrees of freedom or with itself in a manner other than those that are governed by Weyl tensor, which means  no
 derivative occurs to tensors like Ricci, Riemann tensors and Ricci scalar. This suggests the action will be constructed using
the Weyl tensor and we choose the contraction of indices to be of   a very specific kind.

The result of the calculation is that at the fourth power to Weyl tensor, of the type
${{\cal C}^{ij}}_{kl}{{\cal C}^{kl}}_{mn}{{\cal C}^{mn}}_{rs}{{\cal C}^{rs}}_{ij}$, we can see there is an enhancement
to both the shear viscosity and the entropy density and making  the ratio of
$\eta/s$ to increase above $1/4\pi$, provided we take the corresponding coupling $\lambda_4$ to be a positive number.

The paper is organized as follows: in section 2, we shall write down a  low energy effective action in 5-dimension, whose
$\eta/s$ we want to calculate and review briefly the computation of $\eta$ following \cite{liu} and entropy density using
the Wald's prescription \cite{wald}. Then we will present the result of the computation in section 3. We have relegated the explicit computational details of both $\eta$ and entropy density  for two examples, the Gauss-Bonnet term  and square of Weyl tensor  in section 6 of Appendix A and the relation between 
$\lambda_2$ with the central charges in section 7 of Appendix B.

The computation to $\eta/s$ for different examples both with and without the higher derivative corrections have been
studied in \cite{mps}, \cite{bls}, \cite{bhm}, \cite{mmt}, \cite{buchel-liu}, \cite{chls}, \cite{rami}, \cite{bd},
\cite{mas}, \cite{cs}, \cite{ghodsi} and  \cite{others}.

\section{The action and the prescription}
Let us assume
the low energy effective action that we are interested in has the following structure
\bea\label{action}
S&=&\f{1}{16\pi G}\int d^5x\sqrt{-g}\bigg[ R+\f{12}{L^2}+\lambda_2 {{\cal C}^{ij}}_{kl}{{\cal C}^{kl}}_{ij}+
\lambda_3 {{\cal C}^{ij}}_{kl}{{\cal C}^{kl}}_{mn}{{\cal C}^{mn}}_{ij}+\nn &&
\lambda_4 {{\cal C}^{ij}}_{kl}{{\cal C}^{kl}}_{mn}{{\cal C}^{mn}}_{rs}{{\cal C}^{rs}}_{ij}+
\lambda_5 {{\cal C}^{ij}}_{kl}{{\cal C}^{kl}}_{mn}{{\cal C}^{mn}}_{rs} {{\cal C}^{rs}}_{ uv}{{\cal C}^{uv}}_{ij}\bigg],
\eea
where ${\cal C}_{abcd}$ is the Weyl tensor.
In order to evaluate the quantities of interest, we shall follow \cite{liu}, which says that the shear viscosity can
be evaluated by computing the following quantity
\be\label{definition_eta}
\eta=\lim_{k_{a}\rightarrow 0} \f{\prod(r,k_a)}{i\omega \phi(r,k_a)}
\ee
at the boundary, where $\prod$ is the momentum associated to  the field $\phi$ and is related to the  metric fluctuation,
in particular, to the fluctuation in  scalar channel. Let us denote the metric fluctuation
\be
{h^y}_x=\int[dk] \phi_k(r) e^{-i\omega t+i k z},
\ee
where the graviton is moving along  $z$ direction and we are using a short hand notation to write the appropriate measure
factor for  momentum integrals and factors of $2\pi$ in $[dk]$.

Considering this kind of fluctuation, one can show that the field $\phi$ decouples from the rest of the metric
fluctuation \cite{ks}
and its equation of motion can be derived from the following effective action
\bea
S&=&\f{1}{16 \pi G}\int dr [dk]\bigg(A(r)\phi''_k\phi_{-k}+B(r)\phi'_k\phi'_{-k}+C(r)\phi'_k\phi_{-k}+D(r)\phi_k\phi_{-k}+
\nn &&E(r) \phi''_k\phi''_{-k}+F(r)\phi''_k\phi'_{-k} \bigg)+{\cal K},
\eea
where ${\cal K}$ is the appropriately generalized Gibbons-Hawking boundary term, which is written  in \cite{bls},\cite{mps}. One of the important point is that upon inclusion of higher derivative terms to action,  it generically gives
us a non-trivial form of the functions $E(r)$ and $F(r)$. For example, in the 2-derivative action, like the Einstein-Hilbert type, 
these two functions vanish and are non-trivial for other higher derivative terms like Gauss-Bonnet etc. This means these two functions $E(r)$ and $F(r)$ are already of order $\lambda_i$. As we know
via the Kubo formula, the coefficient of shear viscosity  is related to the two point correlation function involving the
energy momentum
tensor and our interest is to compute  it to leading order in the parameters $\lambda_i$, means we shall compute the
functions $A,~B,\cdots,~F,~E$ etc., to linear in $\lambda_i$ and  not like the product of  $\lambda_i$'s or powers of more than unity.

By considering the radial coordinate as time, as is considered in the paper of Liu et  al., we can compute the momentum
and it comes out as
\be
{\prod}_{k}(r)=\f{1}{8\pi G}\bigg[\bigg(B-A-\f{F'}{2}\bigg)\phi'_k(r)-\f{d}{dr}(E\phi''_k) \bigg].
\ee

In order to compute the momentum in  zero frequency limit and then evaluating the ratio eq(\ref{definition_eta}) at
the boundary requires us to know the derivatives of the field $\phi$. In the zero frequency limit the momentum is
constant \cite{liu}, which means it  has the same value either on the boundary or at the horizon. On
evaluating it at the horizon with the in falling
boundary condition for the field $\phi$ requires us to use
\be
\p_r\phi=-i\omega\sqrt{\f{b}{a}}\phi,~~~\p^n_r\phi=-i\omega\bigg(\p^{n-1}_r\sqrt{\f{b}{a}}\bigg)\phi,
\ee
which is argued to work only at the horizon in \cite{liu} and  we have assumed the five-dimensional metric has the
following form, where there is a rotational invariance in the spatial directions
\be\label{metric}
ds^2=-a(r)dt^2+b(r)dr^2+c(r)[dx^2+dy^2+dz^2]
\ee

Using these ingredients that eq(\ref{definition_eta}) gives us, after
dropping the index $k$, of course the momentum dependencies are there in the field, $\phi$ 
\be\label{eta}
\eta=lim_{\omega\rightarrow 0}
\f{\prod}{i\omega\phi}=\f{1}{8\pi
G}[\kappa_2(r)+\kappa_i(r)]_{r=horizon}, \ee where \be
\kappa_2=\sqrt{\f{b}{a}}\bigg[A-B+\f{F'}{2}\bigg],~~~\kappa_i=\f{d}{dr}\bigg[E\f{d}{dr}\sqrt{b/a}\bigg]
\ee

In order to evaluate the ratio $\eta/s$, we need to know the expression to entropy density $s$.
The entropy of the gravitational system can be calculated using  Wald's formula \cite{wald}
\be\label{wald_formula}
{\cal S}=-2\pi\int \bigg(\f{\p L}{\p R_{abcd}}\epsilon_{ab}\epsilon_{cd}\bigg)_{horizon},
\ee
where $L$ is the Lagrangian that  follows from action eq(\ref{action}) and  the binormal quantity $\epsilon_{ab}$ is
normalized to obey  $\epsilon_{ab}\epsilon^{ab}=-2$. One particular choice to construct such an object is to introduce
two null  vectors $\xi_a$ and $\delta_b$ with the restriction $\xi.\delta=1$ on the horizon \cite{dg}
\be
\epsilon_{ab}=\xi_a\delta_b-\delta_a\xi_b,
\ee
with the choice to our metric as written in 
eq(\ref{metric}), it gives
\be
\xi_t=-a,~~~\xi^t=1,~~~\delta_t=1,~~~\delta^t=-\f{1}{a},~~~\delta_r=-\sqrt{\f{b}{a}},~~~\delta^r=-\f{1}{\sqrt{ab}},
\ee
and the rest of the components of $\xi_a$ and $\delta_b$ vanishes.

\section{Computation}

Before doing the computation of $\eta$ for non-zero but small positive $\lambda_i$, let us write down the solution to metric for
the case $\lambda_i=0$. This solution can be read out very easily from the paper \cite{blmsy}  and is given by
\be\label{uncorrected_solution}
a=N^2 \f{(r^4-r^4_0)}{L^2 r^2},~~~b=\f{L^2r^2}{r^4-r^4_0},~~~
c=\f{r^2}{L^2}.
\ee
Let us set $N=1$ or can be re-absorbed in defining the temporal coordinate.
Now, to do the computation for both $\eta$ and ${\cal S}$, we do not need to know the $\lambda_i$ corrected solution\footnote{This we demonstrate by going through two different examples in the appendix. The main point is that the the uncorrected piece to both $\eta$ and entropy density  depends only on the spatial part of the metric component,  which do not receive any correction at the linear order in $\lambda_i$.}.
This is due to the reason that  we are  only interested to calculate these quantities at the leading order in $\lambda_i$ only and 
not in their product. This also means that we can calculate each of the higher derivative terms without the need to worry about
the other terms in the action. It means the action that we need to  consider are the Einstein-Hilbert term, the
cosmological constant term and any one of
the higher derivative term to find the result at that order in derivative expansion. For example, the action that
we shall start out with contains only the $\lambda_2$ term
\be
S_2=\f{1}{16\pi G}\int \sqrt{-g}\bigg[R+\f{12}{L^2}+\lambda_2 ~{{\cal C}^{ij}}_{kl}{{\cal C}^{kl}}_{ij}\bigg]
\ee

Going through the above mentioned procedure as outlined in section
2, this gives us the result that there is an enhancement to both the
coefficient of shear viscosity and to the entropy density ( assuming
$\lambda_i$ is positive, which we do not claim it should always be. ) 

\be\label{eta_weyl_squared} 
\eta= \bigg(\f{r^3_0}{16\pi G
L^3}\bigg) \bigg[1+\f{4\lambda_2}{L^2}\bigg],~~~s\equiv\f{{\cal
S}}{V_3}= \bigg(\f{r^3_0}{4 G L^3}\bigg)
\bigg[1+12\f{\lambda_2}{L^2} \bigg] \ee giving the ratio 
\be
\f{\eta}{s}=\f{1}{4\pi}\bigg[1-8\f{\lambda_2}{L^2}\bigg]. \ee

We do not write down explicitly the form of $\kappa_2$ and
$\kappa_4$ as these are very big expressions and similarly for other cases too. 

 At this order, if we would have considered the Gauss-Bonnet term
instead of the square of the Weyl tensor term then the result on
both the coefficient of shear viscosity and the entropy density
would have been different. In this case there occurs  a suppression
to $\eta$ and unchanged form of entropy density \cite{blmsy}

\be\label{eta_gb}
\eta=\bigg(\f{r^3_0}{16\pi G L^3}\bigg)
\bigg[1-\f{8\lambda_2}{L^2}\bigg],~~~s\equiv\f{{\cal S}}{V_3}=
\bigg(\f{r^3_0}{4 G L^3}\bigg), \ee but the ratio remains the
same \be \f{\eta}{s}=\f{1}{4\pi}\bigg[1-8\f{\lambda_2}{L^2}\bigg],
\ee
as that considered by including the square of Weyl term.\\

Let us move to the next order in derivative expansion and consider the following action
\be
S_3=\f{1}{16\pi G}\int\sqrt{-g}\bigg[ R+\f{12}{L^2}+\lambda_3 {{\cal C}^{ij}}_{kl}{{\cal C}^{kl}}_{mn}{{\cal C}^{mn}}_{ij}\bigg]
\ee

The computation results in the following form to the shear viscosity and the entropy density
\be
\eta=\bigg(\f{r^3_0}{16\pi G L^3}\bigg) \bigg[1-\f{336\lambda_3}{L^4}\bigg],~~~s\equiv\f{{\cal S}}{V_3}=
\bigg(\f{r^3_0}{4 G L^3}\bigg)\bigg[1+48\f{\lambda_3}{L^4} \bigg],
\ee
 and  the ratio
\be
\f{\eta}{s}=\f{1}{4\pi}\bigg[1-384\f{\lambda_3}{L^4}\bigg].
\ee

If we would have taken another structure to the action at this order
instead of considering the cubic power to Weyl tensor, like the one that follows from the third order term to Lovelock gravity
 \bea
S_3&=&\f{1}{16\pi G}\int  \sqrt{-g}\bigg[ R+\f{12}{L^2}+\lambda_3
\bigg(2
R^{ijkl}R_{klmn}{R^{mn}}_{ij}+8{R^{ij}}_{kl}{R^{km}}_{jn}{R^{ln}}_{im}+\nn
&&24R^{ijkl}R_{kljn}{R^n}_i+3RR^{ijkl}R_{klij}+24R^{ijkl}R_{ki}R_{lj}+
16R^{ij}R_{jk}{R^k}_i-\nn &&12R R^{ij}R_{ij}+R^3\bigg)\bigg], \eea
then the shear viscosity is same as that of computing from a 2-derivative  action i.e the Einstein-Hilbert
action. This happens because the momentum associated to the metric fluctuation vanishes, when evaluated at the horizon.   \\

Let us proceed further and do the calculation by including yet another term to action, which is higher than the previous one that is
the fourth power of Weyl tensor. The explicit form the action is
\be
S_4=\f{1}{16\pi G}\int\sqrt{-g}\bigg[ R+\f{12}{L^2}+
\lambda_4 {{\cal C}^{ij}}_{kl}{{\cal C}^{kl}}_{mn}{{\cal C}^{mn}}_{rs}{{\cal C}^{rs}}_{ij}\bigg].
\ee

The computation to shear viscosity and entropy density at this order  results
\be
\eta=\bigg(\f{r^3_0}{16\pi G L^3}\bigg) \bigg[1+\f{1440\lambda_4}{L^6}\bigg],~~~s\equiv\f{{\cal S}}{V_3}=
\bigg(\f{r^3_0}{4 G L^3}\bigg)\bigg[1+480\f{\lambda_4}{L^6} \bigg],
\ee
with the ratio
\be
\f{\eta}{s}=\f{1}{4\pi}\bigg[1+960\f{\lambda_4}{L^6}\bigg].
\ee

So, we see that for the fourth power to Weyl tensor both the
coefficient of shear viscosity and the entropy density increases only for positive $\lambda_4$. Thus making the ratio to increase and respects the KSS bound.

In literature the computation at this order is done by including the other kind of contraction of indices \cite{bls} and the result
is that ratio $\eta/s$ respects the KSS bound. Here we see the result of respecting the KSS bound follows just by considering
only one kind of contraction to indices.

Let us proceed further and include the next order term that is a fifth power of Weyl tensor, with the structure to action
\be
S_5=\f{1}{16\pi G}\int\sqrt{-g}\bigg[ R+\f{12}{L^2}+
\lambda_5 {{\cal C}^{ij}}_{kl}{{\cal C}^{kl}}_{mn}{{\cal C}^{mn}}_{rs} {{\cal C}^{rs}}_{ uv}{{\cal C}^{uv}}_{ij}\bigg]
\ee

The computation results at this order to shear viscosity and entropy density as
\be
\eta=\bigg(\f{r^3_0}{16\pi G L^3}\bigg) \bigg[1-\f{7040\lambda_5}{9L^8}\bigg],~~~s\equiv\f{{\cal S}}{V_3}=
\bigg(\f{r^3_0}{4 G L^3}\bigg)\bigg[1+3200\f{\lambda_5}{L^8} \bigg],
\ee
with the ratio
\be
\f{\eta}{s}=\f{1}{4\pi}\bigg[1-35840\f{\lambda_5}{9L^8}\bigg].
\ee

From the result it just follows that even though there is an
enhancement to entropy density but the suppression to shear
viscosity makes the ratio $\eta/s$ to go below the KSS bound, with the assumption that $\lambda_5$ is positive.

Now we can include all the independent contributions and write the
ratio to $\eta/s$ as \be\label{result_etabys}
\f{\eta}{s}=\f{1}{4\pi}\bigg[1-8\f{\lambda_2}{L^2}-384\f{\lambda_3}{L^4}+986\f{\lambda_4}{L^6}-\f{35840}{9}\f{\lambda_5}{L^8}\bigg],
\ee which we can always do because we are interested to compute the
ratio  only to leading order in $\lambda_i$  and not their product.

We can compare our result with the one that appeared recently
\cite{bhm} and the result  up to the  linear order in $\lambda_3$
matches and the coefficient of $\lambda_4$ and $\lambda_5$ are new.
In the paper \cite{bls}, the authors had done the calculation by
considering a  ${\cal C}^4$ term in the action but the way the
indices are contracted and the number of such terms are different
than we have considered in the present paper. The reason of
considering such a single term at the ${\cal C}^4$ order is just to
show that even with one term we can get a positive contribution to
the ratio $\eta/s$.

From the result of the computation eq(\ref{result_etabys}), it just
follows that to leading in $\lambda_i$, we can rewrite it as \be\label{final_sum}
\eta/s=\f{1}{4\pi}\bigg[1-\sum_{i=2}{\mathcal{\natural}}_i\f{\lambda_i}{L^{2(i-1)}}\bigg],
\ee where ${\natural}_i$ is a constant, which  can take both
positive and negative values but the magnitude of it increases with
the number of derivatives that we are considering in the low energy effective action.  From dimensional analysis, it just
follows that $\lambda_i\sim {\alpha'}^{(i-1)}$, where $\alpha'=l^2_s$, the square of string length, which by AdS/CFT
duality means $\lambda_i\sim \f{1}{\lambda^{\f{(i-1)}{2}}}$, where $\lambda$ is the 't Hooft coupling.

As a check to $\natural_2$ and $\natural_3$, we find that our computation matches with the ones reported in \cite{bhm}.

\section{Conclusion and discussion}
Using the approach of \cite{liu}, we have calculated one of the
transport quantity that is $\eta/s$ and have found some agreement
with the results mentioned in \cite{bhm} and predicted a couple
more. In particular, we have shown that the KSS bound is respected  by considering one
particular type of term at the fourth power to Weyl tensor, provided we take the
coupling $\lambda_4$ as a positive quantity. To the fifth power of
Weyl tensor, the ratio $\eta/s$ do not respect the KSS bound, once
again for positive coupling $\lambda_5$. It is expected on general grounds that 
the couplings $\lambda_i$ in the gravity side are related to the central charges $a$ and $c$  on the dual field theory but the precise relation is not known for all $\lambda_i$, at present. However, $\lambda_2$ is related to $a$ and $c$
via eq(\ref{weyl_squared_coupling})
\be
\f{\lambda_2}{L^2}=\f{1}{8}\f{c-a}{a},
\ee
which are further related to two parameters $t_2$ and $t_4$
defined in \cite{hm} with some constraints on the parameters $t_2$ and $t_4$ coming from the  argument that there should be only 
positive energy that is deposited on the calorimeter "experiment".   
There exists  various restrictions on $a/c$,
depending on the amount of supersymmetry preserved by the system   \cite{hm}. For a system that does not preserve any amount of supersymmetry, one gets the restriction on $\lambda_2$ as $-\f{13}{248}\le \f{\lambda_2}{L^2} \le \f{1}{4}$, for ${\cal N}=1$, it is $-\f{1}{24}\le \f{\lambda_2}{L^2} \le \f{1}{8}$  and for ${\cal N}=2$, it is $-\f{1}{40}\le \f{\lambda_2}{L^2} \le \f{1}{4}$.

Certainly, it is  interesting to  use the  criteria of  
micro-causality violation \cite{blmsy2} to find, if there exists any other constraint on the $\lambda_i$'s. But we do  
not have the explicit gravity solutions for all these cases, which deserve further investigations and possible violation to KSS bound\cite{kp}.

One of the important question that arises in the study is
the convergence of eq(\ref{final_sum}) i.e   how small are the $\lambda_i$'s such that the sum
in eq(\ref{final_sum}) converges ? Is the convergence going to be a criteria to fix the nature of the action other than supersymmetry?

The choice of taking, the kind of action as written in eq(\ref{action}) may not  have been inspired from  string theoretic suggestions,  especially the 8th derivative term. If we take the string theoretic suggestions 
 at this order then  we already know the answer to $\eta/s$ \cite{bls} and the aim here is not to repeat the calculation. 
Rather to show that with a specific kind of contraction to indices at ${\cal C}^4$ order in derivative expansion, we can still
make the KSS bound to hold provided the coupling $\lambda_4$ is positive.  Of course, a priori there is not any reason to accept this kind of low energy effective action. It could be, one may
turn around and ask the question: Is it possible to construct any other type of action at this order in the derivative 
expansion, such that it  respect  KSS bound ? And here is an example.  Also, we are doing phenomenology and trying to find the consequences,  if the low energy effective action has a form the kind eq(\ref{action}).

\section{Acknowledgment}
It is a pleasure to thank Ofer Aharony for going through the manuscript and raising some important questions modulo the computations and the members of  CQUeST for  their help.

This work was supported by the Korea Science and Engineering Foundation (KOSEF) grant funded by the Korea
government (MEST) through the Center for Quantum Spacetime (CQUeST) of Sogang University with grant number R11-2005-021.

\section{Appendix A}

In this appendix, we shall demonstrate  why we do not need to know the exact form of the $\lambda_i$ corrected solution, by going through   two different examples, as stated in the main text. First we shall show for the case, when the gravitational action has two parts: one is Einstein-Hilbert action and the other is of the Gauss-Bonnet kind. Then we shall consider adding the Weyl-squared term. The result is that as long as the spatial part of the metric components  do not receive any corrections to linear in $\lambda_i$, the computation of $\eta/s$ do not require the $\lambda_i$ corrected solution at this order.

\subsection{Einstein-Hilbert and Gauss-Bonnet term}

In this case the action is described by
\be
S=\f{1}{16\pi G}\int d^5x\sqrt{-g}\bigg[ R+\f{12}{L^2}+\lambda_2 (R_{ijkl}R^{ijkl}-4R_{ij}R^{ij}+R^2)\bigg]
\ee  

and it admits an exact solution of the following form
\be\label{lambda_2_corrected_solution}
ds^2=-r^2\alpha^2(r)dt^2+\f{dr^2}{r^2 \beta^2(r)}+\gamma^2(r)(dx^2+dy^2+dz^2).
\ee
We have checked explicitly that this form of the metric solves the equations of motion and 
 the form of the metric components are \cite{blmsy}
\be
\alpha^2(r)=\beta^2(r)=\f{1}{4\lambda_2}\bigg[1-\sqrt{1-8\lambda_2\bigg(1-\f{r^4_0}{r^4}}\bigg)\bigg],~~~\gamma^2(r)=r^2,
\ee
where we have set the size of AdS radius to unity, for convenience. As reviewed in section 2, we can calculate the
coefficient of shear-viscosity using eq(\ref{eta}).  After  completing  the necessary computations to find the expressions for $A,~B,~E$ and $F$, we ended up
\bea
\f{A}{16\pi G}&=&\f{\sqrt{abc^3}}{8\pi G ab^2c^2}[abc^2-\lambda_2 (2  a'cc'+ac'^2)],\nn
\f{B}{16\pi G}&=&\f{\sqrt{abc^3}}{8\pi G a^2b^3c^2}[3 a^2b^2c^2+\lambda_2 (bc^2a'^2+3ac^2a'b'-11abca'c'+3a^2cb'c'-5a^2bc'^2\nn &&-
2abc^2a''-2a^2bcc'')],\nn
\f{F}{16\pi G}&=&-\lambda_2  \f{\sqrt{abc^3}}{8\pi G ab^2c}[ca'+ac'],\nn
\f{E}{16\pi G}&=&0,
\eea

where $a,~b$ and $c$ are the metric components  that appear in  the notation of eq(\ref{metric}). The explicit form of 
\be
\f{\kappa_2}{16\pi G}=\f{c^{3/2}}{32\pi G}-\lambda_2\f{\sqrt{c}a'c'}{32\pi Gab},~~~\kappa_i=0,
\ee
with 
\be
a=b^{-1}=\f{r^2}{4\lambda_2}\bigg[1-\sqrt{1-8\lambda_2\bigg(1-\f{r^4_0}{r^4}}\bigg)\bigg],~~~c=r^2.
\ee

Computing $\eta$ using eq(\ref{eta}) and keeping terms to linear in $\lambda_2$,  gives  the result as
written in eq(\ref{eta_gb}). The main point here is that the term $\kappa_2$ and $\kappa_i$ upon evaluating at the horizon, have the same structure whether we use the $\lambda_2$ corrected solution or not. This essentially says that the extra terms that contribute to $\kappa_2(r)$  vanish upon evaluating at the horizon.

The entropy density calculated using the Wald's formula give the same answer whether we use the $\lambda_2$ corrected solution eq(\ref{lambda_2_corrected_solution}) or
the uncorrected solution eq(\ref{uncorrected_solution}). This is very easy to convince oneself. The reason behind this is that there appears two terms after the differentiation in eq(\ref{wald_formula}), one from differentiating the Einstein-Hilbert term and the other is  from differentiating the Gauss-Bonnet term. The second kind of term is already linear in $\lambda_2$, and our interest is to calculate entropy to  linear order, which means the uncorrected solution eq(\ref{uncorrected_solution}) is good enough for this term. Whereas the first term depends on the spatial part of the  metric components $\gamma^2(r)$, which  do not receive any corrections in $\lambda_2$,  the computation of  entropy density too does not require the $\lambda_2$ corrected solution.

\subsection{Einstein-Hilbert and Weyl-squared term}

In this case the action is described by
\be
S=\f{1}{16\pi G}\int d^5x\sqrt{-g}\bigg[ R+\f{12}{L^2}+\lambda_2 ~{\cal C}_{ijkl}{\cal C}^{ijkl}\bigg]
\ee  

The equations of motion that results 
\bea
&&R_{ij}-\f{1}{2}g_{ij}R-\f{\lambda_2}{2} g_{ij} {\cal W}-6g_{ij}+\lambda_2\bigg[\f{R}{3}R_{ij}+2R_{ikpq}{R_j}^{kpq}
+\f{4}{3}R_{ipjs}R^{ps}-\nn&&4R_{ip}{R_j}^p-\f{1}{2}\nabla_i\nabla_j R-\f{1}{2}\nabla_j\nabla_i R-\f{1}{3}g_{ij}\nabla^2R+\f{4}{3}\nabla^2 R_{ij}\bigg]=0,
\eea
where ${\cal W}={\cal C}_{ijkl}{\cal C}^{ijkl}$ is the 
Weyl-squared term.

This equation of motion admits the following solution
\be
ds^2=-r^2\alpha^2(r)dt^2+\f{dr^2}{r^2 \beta^2(r)}+\gamma^2(r)(dx^2+dy^2+dz^2).
\ee

Generically, it is very difficult to find the exact solution, however for our purpose the solution to linear in $\lambda_2$ is good enough and is 
\be
\alpha=\beta=\f{\sqrt{r^4-r^4_0}}{r^2}\bigg[1-\lambda_2\bigg(\f{r_0}{r}\bigg)^4\bigg]+{\cal O}(\lambda_2)^2,~~~\gamma^2(r)=r^2+{\cal O}(\lambda_2)^2.
\ee

It is interesting to note that $\gamma$ do not receive any correction to linear in $\lambda_2$. Once, again going through the procedure as presented in section 2, gives
\bea
A&=&\f{\sqrt{abc^3}}{48a^4b^4c^4}[96a^4b^3c^4+\lambda_2(-16a^2b^2c^4a'^2-16a^3b c^4a'b'-\nn&&16a^3b^2c^3a'c'+16a^4bc^3b'c'+32a^4b^2c^2c'^2+
32a^3b^2c^4a''-32a^4b^2c^3c'')],\nn
B&=&\f{\sqrt{abc^3}}{48a^4b^4c^4}[72a^4b^3c^4+\lambda_2(12a^2b^2c^4a'^2+12a^3bc^4a'b'+
16a^4c^4b'^2-\nn&&52a^3b^2c^3a'c'-44a^4bc^3b'c'+56a^4b^2c^2c'^2+
8a^3b^2c^4a''-8a^4b^2c^3c'')],\nn
F&=&\lambda_2\f{\sqrt{abc^3}}{48a^4b^4c^4}[-32a^3b^2c^4a'
-64a^4bc^4b'+96a^4b^2c^3c'],\nn
E&=&4\lambda_2\f{\sqrt{abc^3}}{3b^2}.
\eea

Substituting all these terms into eq(\ref{eta}), gives
\be
\eta=\f{c^{3/2}}{16\pi G}+\lambda_2\f{\sqrt{c}}{32\pi G a^2 b^2}[bca'^2+aca'b'-aba'c'-a^2b'c'-2abca''+2a^2bc''],
\ee
which need to be evaluated at the horizon, $r=r_0$. The second piece of this equation is already linear in $\lambda_2$. From the first term, there will not
be any correction at the linear order as $c$ does not receive
any correction at this order. The argument given in the previous subsection  goes through for the calculation of  entropy density and it gives the result as written in eq(\ref{eta_weyl_squared}).

\section{Appendix B: $a$ and $c$ central charges }

Following \cite{no}, we now write down the expressions to central charges $a$ and $c$. For the gravity action 
\be
S=\f{1}{2\kappa^2_5}\int d^5x[R-2\Lambda+\alpha R^2+\beta R^{MN}R_{MN}+\gamma R_{MNKL}R^{MNKL}],
\ee
the central charges are
\bea
\f{c}{16\pi^2}&=& \f{L^3}{\kappa^2_5}\bigg[\f{1}{16}+\f{1}{L^2}\bigg(-\f{5\alpha}{2}-\f{\beta}{2}+\f{\gamma}{4}\bigg)\bigg],\nn
\f{a}{16\pi^2}&=& \f{L^3}{\kappa^2_5}\bigg[\f{1}{16}+\f{1}{L^2}\bigg(-\f{5\alpha}{2}-\f{\beta}{2}-\f{\gamma}{4}\bigg)\bigg],
\eea 
where $L$ is the size of the AdS radius. For a choice like that of the Gauss-Bonnet combination, namely, $\alpha=t_2,~
\beta=-4t_2,~\gamma=t_2$, gives
\be
\f{c}{16\pi^2}=\f{L^3}{16\kappa^2_5}\bigg[1-4\f{t_2}{L^2}\bigg],~~~\f{a}{16\pi^2}=\f{L^3}{16\kappa^2_5}\bigg[1-12\f{t_2}{L^2}\bigg].
\ee

The ratio
\be
\f{a}{c}=\f{1-12\f{t_2}{L^2}}{1-4\f{t_2}{L^2}}=3-\f{2}{1-4\f{t_2}{L^2}}.
\ee

This is the exact result, however, if we want to make contact with eq(5.1) of \cite{blmsy}, then we need to identify $\f{t_2}{L^2}=\lambda_{GB}/2$ and to leading order in $\lambda_{GB}$, 
 
\bea
\f{c}{16\pi^2}&=&\f{L^3}{16\kappa^2_5}\bigg[1-4\f{\lambda_{GB}}{2}\bigg]\simeq \f{L^3}{16\kappa^2_5}\bigg[\sqrt{1-4\lambda_{GB}}\bigg] ,\nn
\f{a}{16\pi^2}&=&\f{L^3}{16\kappa^2_5}\bigg[1-12\f{\lambda_{GB}}{2}\bigg]\simeq \f{L^3}{16\kappa^2_5}\bigg[3~\sqrt{1-4\lambda_{GB}}-2\bigg],\nn
\f{a}{c}&\simeq&3-\f{2}{\sqrt{1-4\lambda_{GB}}}.
\eea

For a choice like that of Weyl-squared combination, $\alpha=\f{t_2}{6},~
\beta=-\f{4}{3}t_2,~\gamma=t_2$, gives
\be
\f{c}{16\pi^2}=\f{L^3}{16\kappa^2_5}\bigg[1+8\f{t_2}{L^2}\bigg],~~~\f{a}{16\pi^2}=\f{L^3}{16\kappa^2_5}.
\ee

It just follows trivially that 
\be\label{weyl_squared_coupling}
\f{t_2}{L^2}=\f{1}{8}\f{c-a}{a}.
\ee

\end{document}